\documentclass{emulateapj}
\usepackage{graphicx}
\usepackage{natbib}

\shortauthors{Singal et al.}
\shorttitle{A Test for Bright Radio Halos}

 \slugcomment{To Appear In ApJL}

\begin{document}

\title{Axial Ratio of Edge-On Spiral Galaxies as a Test For Extended Bright Radio Halos}

\author{J. Singal\altaffilmark{1}, 
A. Kogut\altaffilmark{2}, 
E. Jones\altaffilmark{1}, 
H. Dunlap\altaffilmark{1}}
\altaffiltext{1}{Physics Department, University of Richmond \\
28 Westhampton Way, Richmond, VA 23173}
\altaffiltext{2}{Code 665, NASA Goddard Space Flight Center \\ 
Greenbelt, MD 20771}

\email{jsingal@richmond.edu}

\begin{abstract}
We use surface brightness contour maps of nearby edge-on spiral galaxies to determine whether extended bright radio halos are common.   In particular, we test a recent model of the spatial structure of the diffuse radio continuum by Subrahmanyan and Cowsik which posits that a substantial fraction of the observed high-latitude surface brightness originates from an extended Galactic halo of uniform emissivity.  Measurements of the axial ratio of emission contours within a sample of normal spiral galaxies at 1500 MHz and below show no evidence for such a bright, extended radio halo.  Either the Galaxy is atypical compared to nearby quiescent spirals or the bulk of the observed high-latitude emission does not originate from this type of extended halo.
\end{abstract}


\section{Introduction} \label{intro}

The observed high-latitude radio sky is dominated by a bright diffuse component.  Although the existence of this component  has been recognized for over 30 years \citep{Phillips81, Beuermann85}, its origin remains unclear.  Proposed explanations range from a localized enhancement  in the synchrotron emissivity   \citep{Sun08} to Galactic halo emission \citep{Beuermann85, SC13} to an extragalactic radio background \citep{Fixsen11,Fornengo14}.

The observed diffuse high latitude radio emission is unusual in several respects.  It lacks a counterpart at far-infrared wavelengths,  violating the tight radio/far-IR correlation  observed on scales ranging from as small as tens of pc  to as large as entire galaxies\citep{Condon92,Hughes06,Tabatabaei07}.  An extragalactic origin for the emission would require a previously unobserved population of faint (sub-$\mu$Jy) radio sources \citep{Singal10,Seiffert11,Vernstrom11,Draper11}.  To avoid limits on statistical fluctuations in deep radio surveys, such a population may need to exceed  the number of galaxies in the Universe \citep{Condon12,Ponente11,YL12}.  No such source population has yet been found \citep{Vernstrom14}. On the other hand, a Galactic origin for the emission would require substantial modification to both the cosmic ray and magnetic field distributions \citep{Fornengo14} and would likely overproduce the observed X-ray background via inverse-Compton radiation \citep{Singal10}.   

A recent analysis of multi-frequency radio data by \citet{SC13} (hereafter SC) places the bulk of the observed diffuse high-latitude radio emission into a spherical halo extending $\sim$15 kpc from the Galactic center.  To the extent that the radio halo radius is large compared to the kpc scale height of the diffuse dust emission, the model provides a mechanism to evade the usual radio/far-IR correlation. However, the model predicts that the Galaxy, viewed by an external observer, should show an extended radio halo with surface brightness at a substantial fraction of the Galactic disk.  In this work, we use interferometric radio surface brightness contour maps  of nearby normal\footnote{In this context we use ``normal'' to exclude exceptional galaxies such as LINERs and starbursts which would presumably differ markedly from the Milky Way.} edge-on spiral galaxies to search for a similarly bright and extended halo.

\section{Analysis} \label{analysis}

\subsection{Galactic halo model} \label{halo_model}

\begin{figure}[t]
\centerline{
\includegraphics[width=3.3in,]{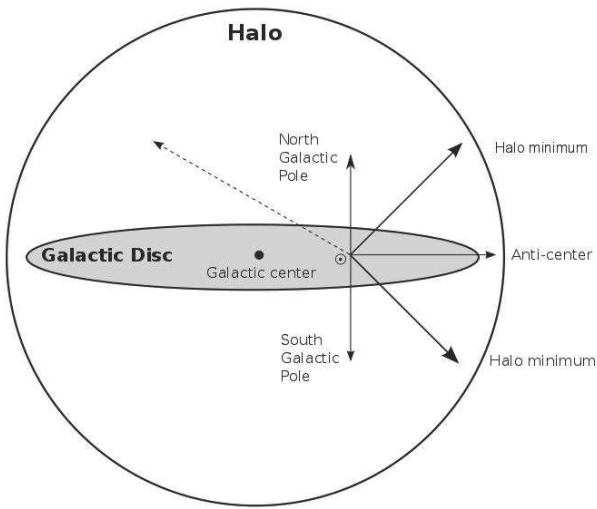}}
\caption{
Disk and halo components from the SC model. The volume emissivities and dimensions of each component are fit to best match 2D maps of the radio sky.  The surface brightness of the halo component is comparable to the disk, and both are assumed to have uniform emissivities per unit volume.  There is an additional component of sources, loops, spurs, and filaments concentrated in the plane. Reproduced from \citet{SC13}.
\label{model_figure} 
}
\end{figure}

Figure \ref{model_figure} shows the relevant features of the SC model.  The diffuse radio sky is fit to a disk component, consisting of an ellipsoid with semi-major axis in the Galactic plane and semi-minor axis toward the Galactic poles, plus a spherical halo component.  The model additionally includes an isotropic extragalactic background as well as an anisotropic contribution from the sum of Galactic sources, loops, spurs, and filaments.

The SC model parameters (extragalactic intensity plus the dimensions and volume emissivities for the ellipsoidal and halo components) are derived by fitting the angular distribution of single dish radio maps at 150, 408, or 1420 MHz.  The resulting halo component contains a substantial fraction of the high-latitude radio brightness.  For example, at 1420 MHz the SC model yields a halo radius 1.8 times the radius of the solar circle, with halo brightness (viewed toward the poles) nearly half that of the disk component viewed in the plane.  The relative importance of the disk and halo components depends on the observing frequency.  The halo component is dominated by synchrotron emission, while the disk component and source components include non-trivial contribution from free-free emission.  We adopt the SC model to search for a similar bright halo among a sample of nearby edge-on spiral galaxies. 

\begin{figure}[b]
\centerline{
\includegraphics[width=3.5in,]{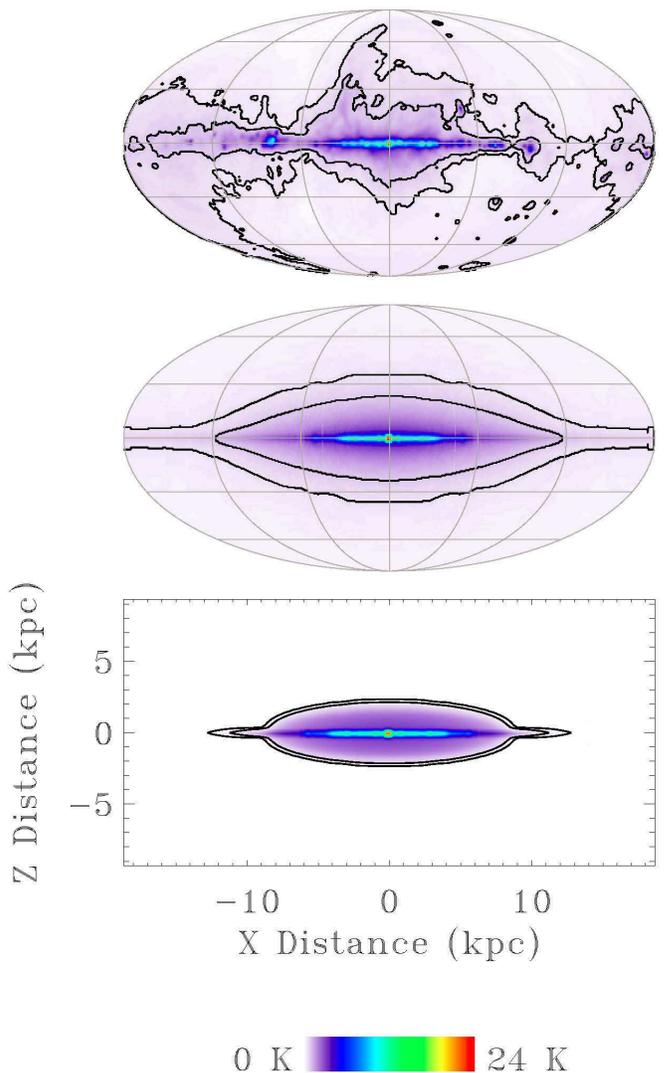}}
\caption{Anisotropic source component of SC model.  (Top) Observed component at 1420 MHz 
after subtracting the SC model diffuse disk, halo, and extragalactic components (Mollweide projection in Galactic coordinates).  The emission is concentrated in the Galactic plane. Contours at 0.4 K and 0.8 K highlight extended emission.  (Middle) Simplified model of anisotropic component as viewed by an observer at the Sun's position in the Galaxy, as discussed in \S \ref{halo_model}. Contours are shown at 0.4 and 0.8 K.  (Bottom) Simplified model of anisotropic component as viewed by an external observer.  Contours are shown at 0.1 K and 0.4 K. All panels use the same linear color bar.
\label{source_fig} 
}
\end{figure}

The dimensions and volume emissivities for the disk and halo components in the SC model specify the large-scale features of the diffuse model emission.  There is also an anisotropic component consisting of sources, loops, and filaments originating from the Galactic plane.  We approximate this component using maps of the radio sky at 1420 \citep{Reich} and 408 MHz \citep{Haslam}.  The top panel of Figure \ref{source_fig} shows the anisotropic component after subtracting the SC model best-fit disk, halo, and extragalactic components from the 1420 MHz survey.  The bulk of the anisotropic emission is concentrated in the Galactic plane.  We model the planar anisotropic emission, as seen by an external observer, using the measured longitudinal distribution as viewed from our vantage point within the Galaxy.  We take the anisotropic contribution for a strip at latitude $|b| = 0$ and longitudes $-60\arcdeg  < l < 60\arcdeg$ then add the contribution in the opposite direction ($l + 180\arcdeg$) to get the combined emission for lines of sight interior to the solar circle.  We then convert the longitudinal profile to linear distance from the Galactic center and extrapolate the resulting profile to match the major axis length of the diffuse disk component, assuming uniform volume emissivity beyond the solar circle.

Although the bulk of the anisotropic emission is concentrated in the Galactic plane, fainter features extend to higher latitude, with typical amplitude 0.5 K at latitude $|b| = 30\arcdeg$.  We model the combined contribution of faint extended features as an additional ellipsoidal component of uniform emissivity to add to the anisotropic component in the plane to achieve a full model of the source component.  We adopt a major axis length equal to the length of the planar component and conservatively adopt a scale height of 2 kpc to maximize the potential confusion of this component with a bright spherical halo.  The bottom panel of Figure \ref{source_fig} shows the resulting combined model of anisotropic emission as viewed by an external observer.  We can scale the anisotropic component brightness in frequency by a position-dependent spectral index derived by comparing this component in maps at 408 and 1420 MHz, with typical value -2.5.  Since we search for evidence of a halo extending well beyond a few kpc, our results are insensitive to the details of the anisotropic component (see the discussion in $\S$3).

\begin{figure*}[t]
\centerline{
\includegraphics[width=7.0in,]{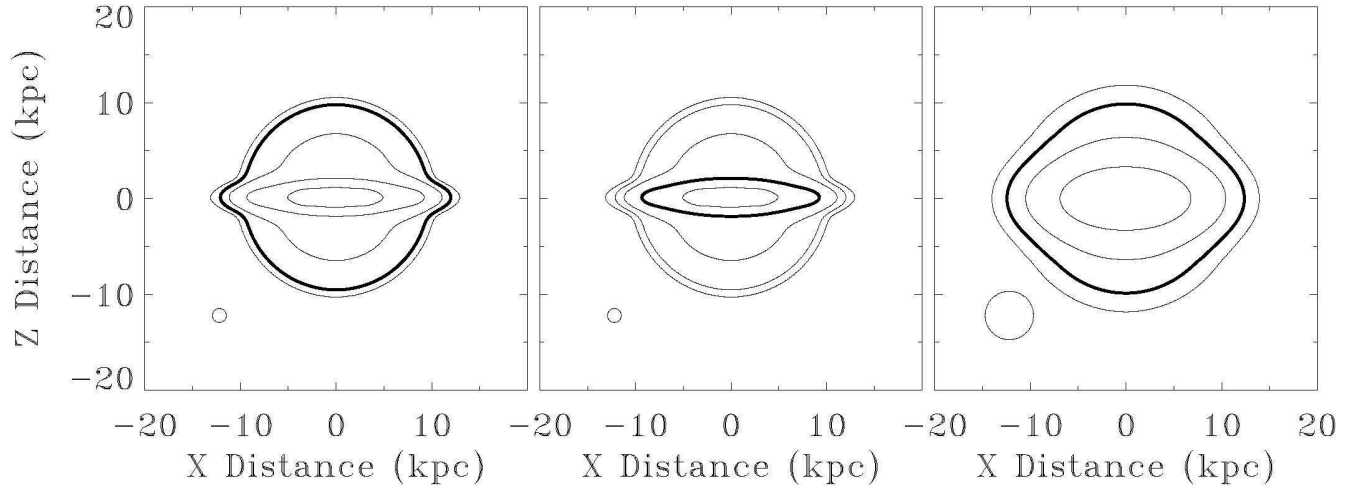}}
\caption{
Predicted radio surface brightness contours for the SC disk$+$halo$+$source model, showing the effects of beam smearing. Contours are shown at 0.03, 0.1, 0.3, 1, and 3 K as seen by an external observer at 1500 MHz for different beam widths. The beam full width at half maximum (FWHM) is shown at the bottom left in each panel.  (Left) Model smoothed with projected beam width of 1.5 kpc.  The plane and halo components are readily distinguished.  The thick contour at 0.1 K highlights the halo component.  (Middle) Same as left panel, but with the thick contour at 1 K highlighting that at this level the contour is dominated by plane component.  Contours near peak brightness are insensitive to the halo component regardless of beam smearing.  (Right) Model smoothed with projected beam width of 5 kpc with thick contour highlighted at 0.1 K.  Beam smearing in this case prevents identification of the halo component even at faint contours.
\label{model_contours} 
}
\end{figure*}

Figure \ref{model_contours} shows the predicted brightness distribution for the full SC disk$+$halo$+$source model at 1500 MHz, as viewed edge-on by an external observer.  Both the dominant plane and fainter halo components are apparent.  Contours near the peak surface brightness trace the plane component, while contours at lower surface brightness have the potential to trace the halo component.  The ability of a contour to distinguish between the two components depends on both angular resolution and surface brightness sensitivity.  Observations at low surface brightness for which beam smearing is not too severe are potentially sensitive to the halo component, while those at higher surface brightness or larger beam width are dominated by the plane component.

The transition from elongated contours near the peak surface brightness to circular contours at lower surface brightness suggests a test for halo emission.  We select a contour from a radio map of an external edge-on spiral galaxy and measure the major axis $A$ and the minor axis $B$ for that contour.  The axial ratio $A/B - 1$ is a measure of the ellipticity, with large values corresponding to highly elliptical distributions and low values tending toward circular.  We then compare the observed axial ratio to the prediction from the model given the observed contour level (specified either as an absolute brightness temperature or a fraction of the peak surface brightness), inclination, and beam width.

\subsection{Radio contour maps}  \label{mapsub}

We use radio surveys of nearby edge-on spiral galaxies to determine the relative abundance of extended bright radio halos.  We restrict analysis to nearby galaxies to enable mapping with sufficient angular resolution to distinguish a spherical halo (see, e.g., the discussion in \citet{Irwin00}), and to galaxies with inclination angle $\theta > 75\arcdeg$ to reduce confusion between the disk and halo components.  The relative importance of the disk and halo components depends on the observing frequency.  We thus consider only galaxies mapped at frequencies between 150 and 1500 MHz to minimize scaling uncertainties and to concentrate on frequencies where the halo is most prominent.

We consider radio contour maps of edge-on spiral galaxies at 1400-1500 MHz provided in \citet{Irwin99}, \citet{HV89}, \citet{Elmouttie97}, \citet{Hummel88}, \citet{Beck94}, \citet{O07}, and \citet{Dahlem97}, along with those at 1400 MHz and 330 MHz in \citet{Carilli92}, 1407 MHz and 408 MHz in \citet{Pooley69}, and 1465 MHz and 308 MHz in \citet{Sukumar88}.  We additionally consider 16 of the 18 galaxies with $\theta > 75\arcdeg$ mapped at 1490 MHz by \citet{Condon87}.  The two we exclude are M82 and M104, which are an irregular starburst and a LINER, respectively, and therefore not representative of Milky Way-like spirals.  This leads to a total of 34 nearly edge-on spiral galaxies mapped at or near 1500 MHz, of which three are also mapped at or near 408 MHz, and of which ten are mapped twice at or near 1500 MHz, for a total of 47 individual maps.  As discussed in $\S$\ref{compsec}, only a subset of these maps have the sensitivity to distinguish the presence or absence of an extended halo similar to that in the SC model.

Interferometric observations are insensitive to structure on angular scales larger than a certain limit, corresponding to the length scale of the shortest baseline.  The resulting loss of sensitivity to extended structures can mask the presence of radio halos.  For the present analysis we selected those observations from the literature where this would not be a significant limitation.  \citet{Irwin99}, \citet{Condon87}, \citet{HV89}, \citet{Hummel88}, \citet{Beck94}, and \citet{Dahlem97}, which together provide the majority of contour maps used in this analysis, use Very Large Array (VLA) observations in C or D configuration.  The shortest baseline is 35 m, corresponding to angular scales of 970\arcsec ~at 1500 MHz.  With the exception of one galaxy, the largest angular scale in the VLA data near 1500 MHz used here is 836\arcsec, within the VLA's resolving capabilities.  In the case of maps of NGC 0253 the major axis angular size exceeds this scale but the minor axis does not, so these maps may still be sensitive to the presence of a spherical halo, and the halo structure seen in these maps is consistent with combined interferometric and single-dish data at 5 GHz \citep{Heesen09}.  The shortest baseline in the Australia National Telescope Facility's Compact Array, used in the observations of \citet{Elmouttie97}, is 30.6 m, enabling observations well beyond the 685\arcsec~used in this work.  The Westerbork Synthesis Radio Telescope, used in the observations of \citet{O07}, has a shortest baseline of 36 m, with a largest observable structure of 840\arcsec~at 1420 MHz, which is larger than the 700\arcsec ~maximum extent of the largest contour in that work.  The Ooty Synthesis Radio Telescope, used for the maps in \citet{Sukumar88}, has a largest observable structure of 40\arcmin~\citep{Ooty88}.  Lastly, the Cambridge One Mile Telescope used for the observations in \citet{Pooley69} has an adjustable baseline.  The precise configuration used in those observations is not readily available in the literature, making it difficult to estimate the maximum scale of resolvable structure for the one galaxy from that work.

\subsection{Axial ratio comparison} \label{compsec}

We test for the presence of a bright radio halo by measuring the axial ratio $A/B - 1$ of edge-on spiral galaxies. We fit the SC model parameters (including the anisotropic source component) to individual power laws in frequency to scale the model to the observed frequency\footnote{Since all but one of the maps are observed at frequencies very close to the frequencies 408 MHz or 1420 MHz used by the SC model, the uncertainty in the scaling is minimal.}.

We then create a contour map of the resulting model (including the effects of galaxy inclination and observing beam width) and measure the axial ratio of the model at the same contour as the faintest contour in the radio data, thereby providing a direct comparison of the model to observations.

Not all of the radio maps are sensitive to extended diffuse emission, although in all maps the faintest contour is the contour most likely to be sensitive.  As seen in Figure \ref{model_contours}, when the faintest contour is not at a sufficiently low surface brightness it is dominated by emission in the plane regardless of the presence or absence of a radio halo.  In maps with significant beam smearing, even the faintest contour cannot distinguish between (convolved) planar emission and diffuse extra-planar emission.  For each radio map, we generate the corresponding model map with and without the halo component, and compare the axial ratios of the resulting models evaluated at the faintest contour level of the radio map.  Radio maps for which the faintest contour fails to distinguish between the presence or absence of the SC model halo component are not used in this analysis.

\begin{figure}[t]
\centerline{
\includegraphics[width=3.5in,]{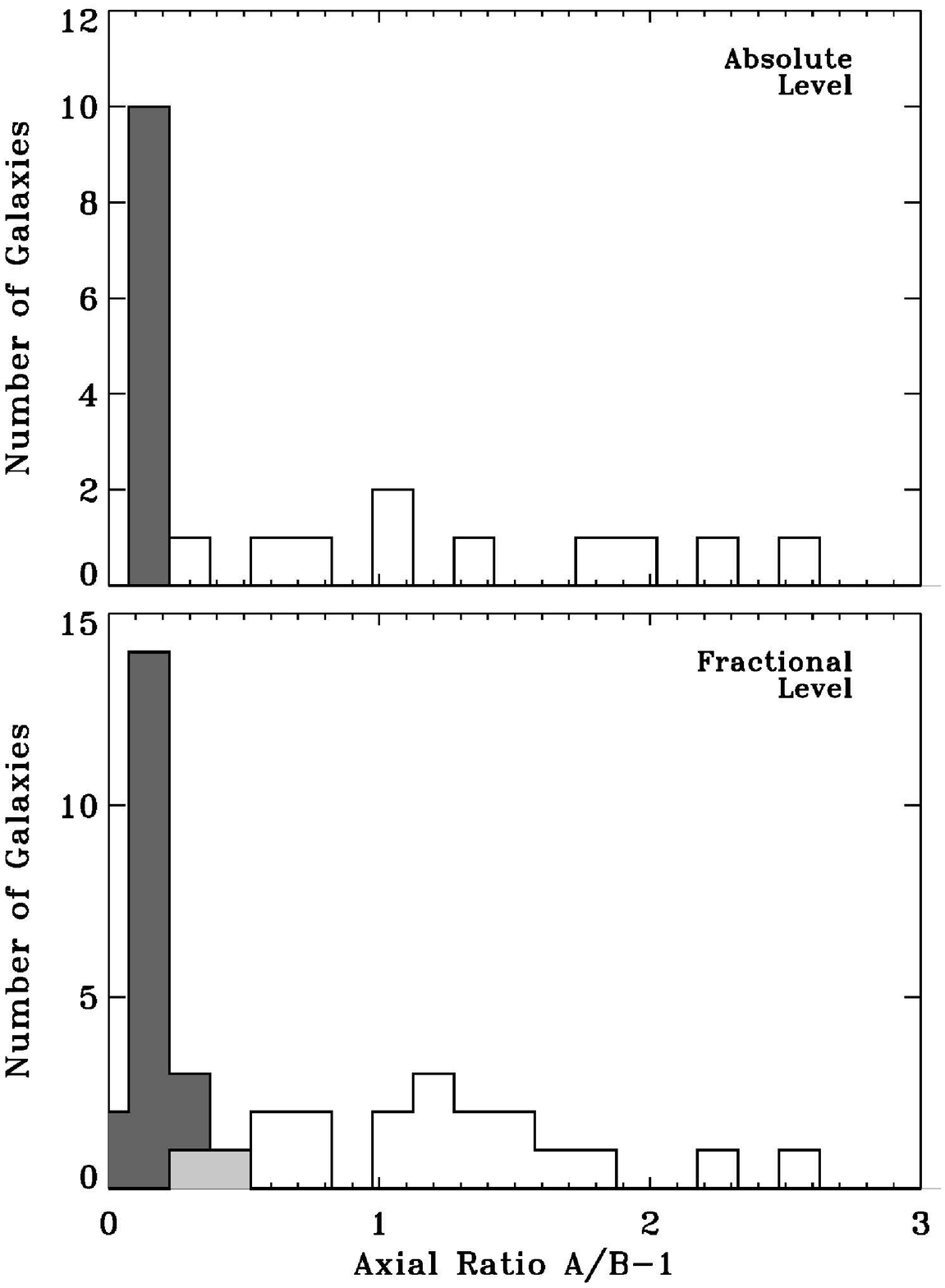}}
\caption{Histogram of measured axial ratios for external edge-on spiral galaxies, for those galaxies with map contours capable of distinguishing the presence of an SC model halo.  The top panel compares the faintest contour from each radio map to the same absolute contour in the model. The bottom panel compares the faintest contour from each radio map to the model contour at the same fraction of the peak surface brightness.  The white histogram shows the observed radio data while the dark histogram shows the model prediction, taking into account the observing frequency, inclination angle, and beam smearing.  Grey cells are common to both data and model histograms.  Edge-on spiral galaxies fail to show the bright halo component predicted by the SC model.
\label{axial_hist} 
}
\end{figure}

We would like to search for extended halos of the type predicted by the SC model in a way that is not biased by galaxies' differing absolute luminosities or the details of planar emission.  We test for these biases based on galaxy luminosity by repeating the axial ratio analysis using two different methods.  The first method uses the absolute surface brightness of the faintest contour in each radio map, and compares the observed axial ratio of that contour to the same absolute contour in the model.  Since the halo emission extends far from the plane, this method is insensitive to details of emission near the plane, but assumes consistent halo brightness for all galaxies.  The second method scales the contours by the peak surface brightness, comparing the faintest contour in each radio map to the same fractional contour in the model.  This method allows the halo brightness to vary from one galaxy to another, but introduces some potential dependence on the details of peak emission in the plane.

Both methods yield similar results (Table 1).  Ten of the radio maps have absolute contours comparable to the predicted halo surface brightness ($\sim$0.1 K at 1500 MHz) and have beam width small compared to the galaxy major axis.  Allowing the halo temperature to scale with the observed peak surface brightness yields 20 maps with a contour capable of detecting a scaled radio halo.  Figure \ref{axial_hist} shows the distribution of observed and model axial ratios.  After accounting for the spread in inclination angle, beam smearing, and observing frequency, the SC model predicts axial ratios $0.23 \pm 0.02$ for the faintest absolute contours and $0.24 \pm 0.09$ for the faintest fractional contours, consistent with the expected spherical halo component.  The radio data, in contrast, show strongly elliptical contours, with mean axial ratios $1.4 \pm 0.7$ and $1.5 \pm 1.1$ for the absolute and fractional contours, respectively.  All of the observed galaxies had an axial ratio larger than the corresponding model prediction.

\section{Discussion}

22 of the 47 radio maps, sampling 13 unique galaxies, had both the sensitivity and angular resolution to test for a bright extended halo of the type predicted by the SC model under one or both comparison methods.  All of the these maps show elongated contours at surface brightness values where a halo should be detectable, while the model consistently predicts rounder contours.  We now consider modifications to the model that would be required to bring the model axial ratio closer to the observed values.  The SC model includes emission from discrete sources, spurs, and radio loops, which we approximate using the Galaxy as a template.  Similar emission is observed in external galaxies and is largely confined to the plane (see, e.g., the discussion in \citet{Irwin99, Irwin00, HV89}).  Since this emission does not extend to the $\sim$15 kpc scale heights of the SC halo, contours using the absolute surface brightness which can distinguish the presence of a halo are not affected by this component.  Relative contours, though, can be affected if the model systematically underestimates the peak surface brightness of the source component in the galaxies, thereby biasing the model relative contours to fainter (and therefore rounder) values than the corresponding galaxies.  Increasing the model axial ratio to the mean value observed in external galaxies requires increasing the surface brightness of the source component by a factor of five. Alternatively, we can increase the model ellipticity by retaining the nominal surface brightness of the source component, but extending the major axis of this component well beyond that of the diffuse disk component.  In this case, matching the model axial ratio to the radio data requires a source component with linear extent over twice that of the diffuse disk component.

\begin{table*}[t]
\begin{center}
\caption{Measured and model axial ratios} 
\label{results_table}
\begin{tabular}{c c c c c c c c l}
\tableline
\multicolumn{9}{c}{Absolute Contour Level Comparison}\\
Galaxy	& Frequency & Inclination & FWHM      & Beam       & Contour & \multicolumn{2}{c}{Axial Ratio $A/B - 1$} & Reference \\
	& (MHz)     & (Deg)       & (\arcsec) & Ratio$^a$  &   (K)   & Data   & Model \\
\tableline
\\
NGC 0253 &  1500 &  78  &  74.0 &  17.8  &  0.11   &   1.47  &       0.20         &  \cite{Beck94} \\
NGC 0891 &  1420 &  89  &  14.7 &  47.7  &  0.10   &   1.06  &	  0.23	&  \cite{O07} \\
NGC 0891 &  1490 &  89  &  48.0 &  13.2  &  0.12   &   1.16  &	  0.25	&  \cite{Condon87} \\
NGC 3628 &  1490 &  87  &  54.0 &  12.5  &  0.09   &   1.90  &	  0.22	&  \cite{Condon87} \\
NGC 4192 &  1490 &  83  &  54.0 &   7.6  &  0.07   &   0.88  &	  0.20	&  \cite{Condon87} \\
NGC 4216 &  1490 &  89  &  54.0 &   6.8  &  0.09   &   1.99  &	  0.23	&  \cite{Condon87} \\
NGC 4565 &  1490 &  86  &  60.0 &  11.6  &  0.07   &   2.59  &	  0.21	&  \cite{Condon87} \\
NGC 4631 &  1490 &  85  &  60.0 &  13.8  &  0.15   &   0.69  &	  0.25	&  \cite{Condon87} \\
NGC 4631 &  1490 &  85  &  40.0 &  20.9  &  0.16   &   0.30  &        0.25       &  \cite{Hummel88} \\
NGC 5907 &  1490 &  88  &  48.0 &  10.8  &  0.08   &   2.35  &	  0.21	&  \cite{Condon87} \\
\\
\tableline
\multicolumn{9}{c}{Fractional Contour Level Comparison}\\
Galaxy	& Frequency & Inclination & FWHM      & Beam       & Contour & \multicolumn{2}{c}{Axial Ratio $A/B - 1$} & Reference  \\
	& (MHz)     & (Deg)       & (\arcsec) & Ratio$^a$  &   (\%)   & Data   & Model \\
\tableline
\\
NGC 0253 &   330  & 78  &  30.0  & 41.8   &  0.5   &   0.47  &         0.04        &  \cite{Carilli92} \\
NGC 0253 &  1400  & 78  &  30.0  & 64.5   &  0.02   &   1.32  &	  0.36	&  \cite{Carilli92} \\
NGC 0253 &  1500  & 78  &  74.0  & 17.8   &  0.4   &   1.47  &	  0.21	&  \cite{Beck94} \\
NGC 0891 &  1490  & 89  &  48.0  & 13.2   &  0.6   &   1.16  &	  0.22	&  \cite{Condon87} \\
NGC 2613 &  1500  & 79  &  16.0  & 20.4   &  3.3   &   1.51  &	  0.31	&  \cite{Irwin99} \\
NGC 3044 &  1465  & 86  &  15.0  & 17.3   &  1.6   &   1.22  &	  0.22	&  \cite{HV89} \\
NGC 3044 &  1500  & 86  &  12.2  & 15.8   &  2.5   &   1.07  &	  0.28	&  \cite{Irwin99} \\
NGC 3221 &  1465  & 84  &  15.0  & 11.9   &  1.6   &   1.77  &	  0.21	&  \cite{HV89} \\
NGC 3221 &  1500  & 84  &  11.7  &  9.2   &  1.0   &   1.59  &	  0.20	&  \cite{Irwin99} \\
NGC 3556 &  1500  & 75  &  15.6  & 27.9   &  1.3   &   1.22  &	  0.18	&  \cite{Irwin99} \\
NGC 3628 &  1490  & 87  &  54.0  & 12.5   &  0.2   &   1.90  &	  0.23	&  \cite{Condon87} \\
NGC 4157 &  1465  & 82  &  14.4  & 16.8   &  2.5   &   0.85  &	  0.27	&  \cite{Irwin99} \\
NGC 4157 &  1465  & 82  &  15.0  & 17.5   &  3.1   &   0.75  &	  0.29	&  \cite{HV89} \\
NGC 4192 &  1490  & 83  &  54.0  &  7.6   &  2.2   &   0.88  &         0.22        &  \cite{Condon87} \\
NGC 4517 &  1500  & 85  &  13.5  & 32.4   &  2.5   &   5.65  &        0.49        &  \cite{Irwin99} \\
NGC 4565 &  1490  & 86  &  60.0  & 11.6   &  3.1   &   2.59  &	  0.30	&  \cite{Condon87} \\
NGC 4631 &   408  & 85  &  96.4  & 13.8   &  5.6   &   1.38  &	  0.04	&  \cite{Pooley69} \\
NGC 4631 &  1490  & 85  &  40.0  & 20.9   &  0.4   &   0.30  &	  0.23	&  \cite{Hummel88} \\
NGC 4631 &  1490  & 85  &  60.0  & 13.8   &  0.6   &   0.69  &	  0.22	&  \cite{Condon87} \\
NGC 5907 &  1490  & 88  &  48.0  & 10.8   &  2.2   &   2.35  &	  0.24	&  \cite{Condon87} \\
\\
\tableline
\multicolumn{9}{l}{$^a$Ratio of major axis to beam FWHM. 
Maps with beam ratio $<$6 confuse halo with plane emission and are not used.} \\
\end{tabular}
\end{center}
\end{table*}

The failure of the SC model to predict the axial ratios for nearby edge-on spirals suggests that this type of bright, extended radio halo can not be a common feature of quiescent spiral galaxies.  The SC model halo in combination with the other model components reproduces the observed diffuse radio sky and the parameters are derived by fitting the angular distribution of single dish radio maps.   Thus, while an extended halo of uniform emissivity is sufficient to model the observed high latitude diffuse radio emission if it is assumed to be predominantly Galactic in origin, it is not the only possible type of extended halo.  Indeed higher resolution observations seem to point to exponential emission profiles in halos \citep[e.g.][]{DK98}, although these halos have characteristic scale heights of only a few kpc above the plane.

We may place a limit on the contribution of the type of extended radio halo posited by the SC model by adjusting the volume emissivity of the halo component in the SC model to match the axial ratio distribution observed in external edge-on spirals.  With the halo radius fixed, the volume emissivity of the halo component must be decreased by a factor $0.27 \pm 0.22$ to match the mean axial ratio $1.4 \pm 0.7$ observed for the sample of nearby normal spirals.  A halo component at this reduced level is compatible with observations of external galaxies, but can no longer fully explain the isotropic component of the radio sky.  The difference between the original halo component in the SC model and the reduced amplitude derived from comparison with external galaxies amounts to 5 K at 408 MHz, as viewed from the Sun's location in the Galaxy.  If the difference is added to the isotopic extragalactic background component of the SC model, the background component has a revised amplitude of $10 \pm 1$ K (not including the 2.7 K cosmic microwave background).  A radio background at this amplitude is consistent with independent estimates $11.8 \pm 1.1$ of the radio background \citep{Fixsen11,Fornengo14}.

We conclude that measurements of nearby edge-on spiral galaxies show no evidence for a bright radio halo comparable to the halo proposed by \citet{SC13}.  Either the Galaxy is atypical compared to nearby quiescent spirals or the bulk of the observed high-latitude Galactic emission does not originate from such an extended halo.

\acknowledgments

Support for this research comes from NASA's Science Innovation Fund.

\clearpage

\end{document}